\documentclass[%
reprint,
onecolumn,
showpacs,
preprintnumbers,
amsmath,amssymb,
pra,
superscriptaddress
]{revtex4-1}
\usepackage{graphicx,subfigure}% Include figure files
\usepackage{dcolumn}% Align table columns on decimal point
\usepackage{bm}% bold math
\usepackage{float}
\usepackage{color}
\usepackage[all]{nowidow}
\usepackage{graphicx}
\usepackage{amsmath}
\usepackage{amsfonts} 
\usepackage{microtype}
\begin{document}

\title{Tunable vortex dynamics in proximity junction arrays: a possible accurate and sensitive 2D THz detector}
\author{J. Rezvani}
\affiliation{Istituto Nazionale di Fisica Nucleare - Laboratori Nazionali di Fisica Nucleare, Via Enrico Fermi, 00044 Frascati, Italy.}
\affiliation{Consiglio Nazionale delle Ricerche (CNR), CNR-IOM, 34149 Basovizza, Italy.}
\author{D. Di Gioacchino}
\affiliation{Istituto Nazionale di Fisica Nucleare - Laboratori Nazionali di Fisica Nucleare, Via Enrico Fermi, 00044 Frascati, Italy.}
\author{C. Gatti}
\affiliation{Istituto Nazionale di Fisica Nucleare - Laboratori Nazionali di Fisica Nucleare, Via Enrico Fermi, 00044 Frascati, Italy.}
\author{N. Poccia}
\affiliation{Institute for Metallic Materials, IFW-Dresden, Dresden, 01069, Germany.}
\author{C. Ligi}
\affiliation{Istituto Nazionale di Fisica Nucleare - Laboratori Nazionali di Fisica Nucleare, Via Enrico Fermi, 00044 Frascati, Italy.}
\author{S. Tocci}
\affiliation{Istituto Nazionale di Fisica Nucleare - Laboratori Nazionali di Fisica Nucleare, Via Enrico Fermi, 00044 Frascati, Italy.}
\author{M. Cestelli Guidi}
\affiliation{Istituto Nazionale di Fisica Nucleare - Laboratori Nazionali di Fisica Nucleare, Via Enrico Fermi, 00044 Frascati, Italy.}
\author{S. Cibella}
\affiliation{Consiglio Nazionale delle Ricerche, Istituto di Fotonica e Nanotecnologie, Rome, 00156, Italy.}
\author{S. Lupi}
\affiliation{Department of Physics, Università degli Studi di Sapienza, 00185 Rome, Italy.}
\author{A. Marcelli}
\affiliation{Istituto Nazionale di Fisica Nucleare - Laboratori Nazionali di Fisica Nucleare, Via Enrico Fermi, 00044 Frascati, Italy.}
\begin{abstract}
An array of superconducting proximity islands has been shown to be highly tunable by electric and magnetic fields. Indeed, a small change in the electric and magnetic field can tune the system from a vortex Mott insulator to a vortex metal. This transition from localized to a non-localized state can occur as soon as the density of the superconducting vortices matches the density of the pinning sites in a non defective structure. The possibility of further modulation of non-localized superconducting states via enhancement of the superconducting order parameter or modulation of the Josephson plasma frequency is discussed. Based on the non-bolometric effects associated with the occurrence of non-equilibrium phenomena in this original superconducting networks we discuss also the possible applications of this array for a conceptually new type of radiation detector. 
\end{abstract}
\maketitle 

\section{Introduction}
The terahertz (THz) region is often described as the final unexplored area of the electromagnetic spectrum. In this energy region, different physical phenomena such as the phonon dynamics and the bound states of quantum wells can be investigated. However, THz instruments and detection capabilities are limited in comparison with the adjacent IR domain. Hence, construction of novel sources and detectors with enhanced capabilities open possibilities in many fields from the solid state (phonon dynamics), astrophysics (galactical formation and evolution) to particle physics (axions and cold dark matter) investigations \cite{Jepsen2011,Saeedkia2013,Karasik2011,rez}. Semi and Superconducting materials are known to show extraordinary properties at low dimensions \cite{rezge,Rezvani_2016,Pinto2018}. An array of superconducting proximity islands has been shown to be highly tunable by electric, thermal and magnetic fields \cite{Poccia2015,Lankhorst2018}. Indeed, a small change in the electric, thermal and magnetic field can tune the array from a vortex Mott insulator to a vortex metal.  A vortex Mott insulator occurs in a type II superconductor if the density of the superconducting vortices matches the density of the pinning sites \cite{Lankhorst2018,goldberg} however only in the array of superconducting island, free from disorder contributions, the dynamic state has been unveiled. We will discuss in this contribution, further superconducting dynamic properties of non-localized vortices and the possible applications of such array for a conceptually new type of radiation detector with an extremely high-energy resolution. This device is based on the non-bolometric effects associated with the occurrence of non-equilibrium phenomena in superconducting networks such as universal scaling properties of the current and magnetic field of the vortex Mott insulator dynamic transition and the modulation of the superconducting order parameter.
\section{experiment}
The system we designed and realized consists of an array of about 90000 Nb superconducting islands regularly deposited on a non-superconducting support of $80\times80$ $\mu m^2$ in size and with a period of $\sim270$ nm, with an island diameter of $\sim220$ nm and a thickness of 45 nm (see Fig.\ref{Tc}). Standard photo-lithographic tools have been employed to obtain a 40 nm thick Au template on a Si/SiO$_2$ substrate of $1\times1$ cm$^2$ as a non superconducting layer. The template consists of a central square of $82\times82$ $\mu$ m$^2$, with the corners connected to 4 terminals for electric measurements. The size of the terminals is $200\times200$ $\mu$ m$^2$, large enough for micro bonding. The Nb pattern is then created on top of the central Au square employing electron beam lithography and DC sputtering. To ensure a uniform current injection into the array, two 45 nm Nb bus bars are deposited along two opposite sides of the array. The IV measurements are carried out in a shielded cryostat at 4.2 K. A current bias is applied using a ramp generator at 107 Hz. Two measurements mode were used. In one the I-V curves were measured amplifying the output voltage via low noise amplifier and digitized by the National instruments sampling card. In second mode, the voltage sinusoidal waveform was measured via a lock-in amplifier locked to the applied current signal resulting in a direct measurement of the dynamic resistance. A magnetic field is applied by placing a solenoid around the sample with the field perpendicular to the sample plane. The current through the solenoid increased in a stepwise fashion and separate I-V traces were recorded at each field step.
\section{Results and discussions}
\begin{figure}[h!]
\includegraphics[width=0.4\textwidth]{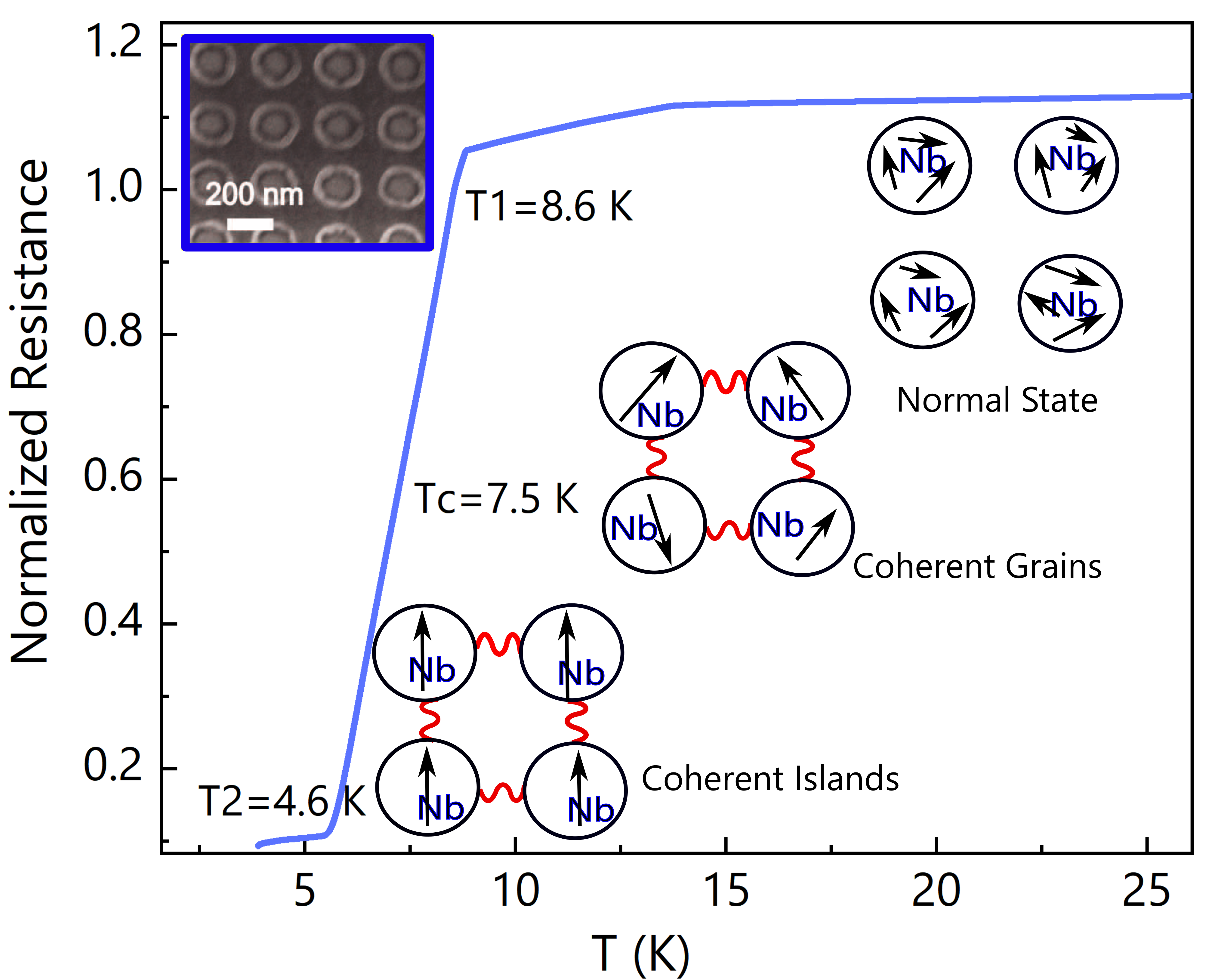}
\caption{Multi stage transition of the Nb islands on the Au layer. Island transition is marked at T$_1$ and the film transition marked at T$_2$. Above T$_1$, the Nb islands are normal metals. Arrows indicate the coherence in grains and islands. Below T$_1$ grains start to become coherent throughout each island. At T$_1$,Cooper pairs start their diffusion from the Nb into the Au, and the resistance drops. Gradually the coherence length of the film becomes comparable to the island spacing, and the entire system progresses toward a global phase coherence. As the temperature decreases (T$_2$), the film undergoes a transition to a superconducting state. (Inset): The SEM images of the Nb islands on Au film.}
\label{Tc}
\end{figure}

\begin{figure}
\centering
\includegraphics[width=0.4\textwidth]{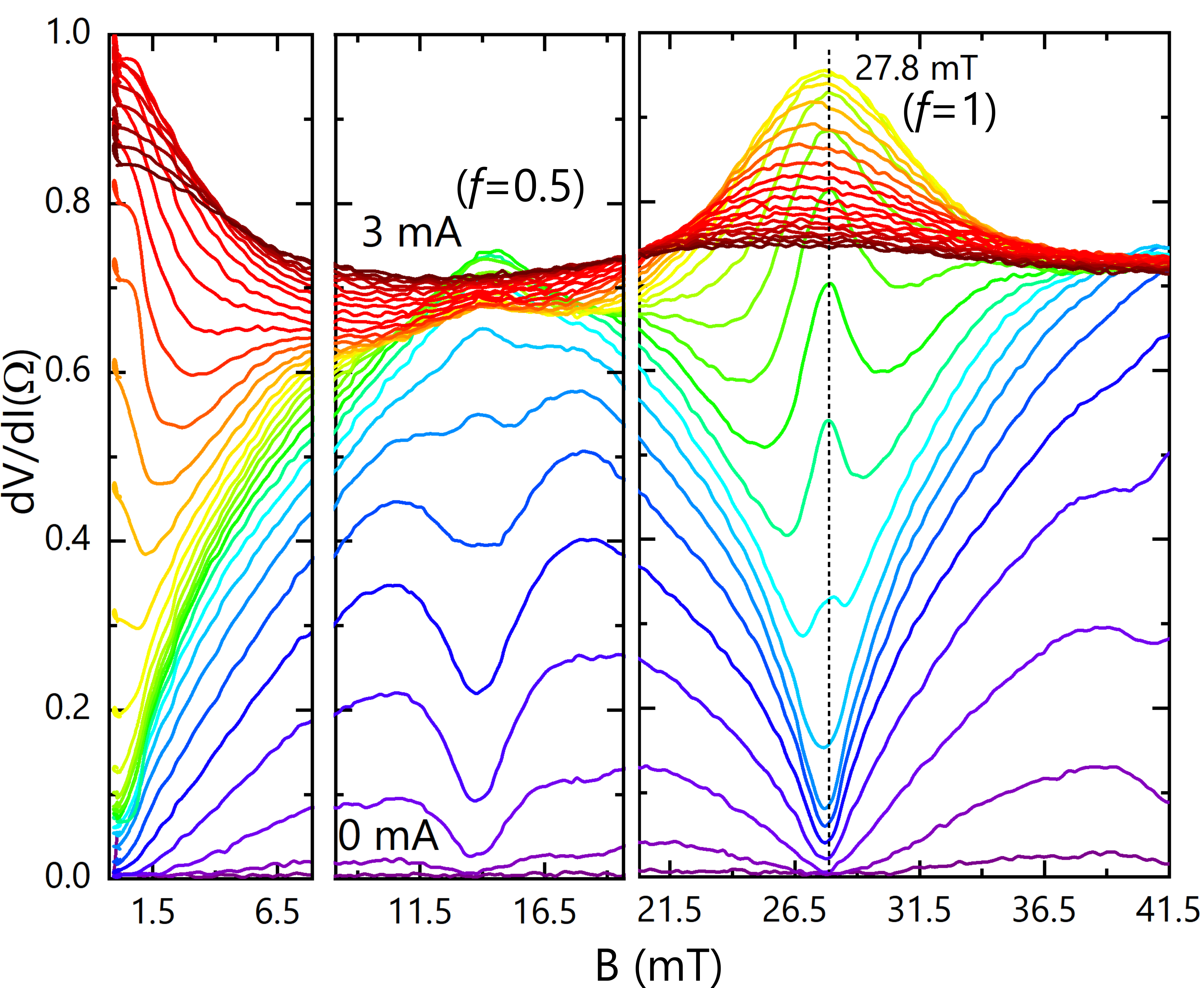}
\caption{Representative dV/dI versus magnetic field curves at different bias currents. At low current bias, dV/dI minima at different B values indicate formation of a vortex Mott insulator. Increasing current reverses minima into maxima manifesting the insulator-to-metal transition. The The transitions are most pronounced at integer and half-integer frustration factors ($f=1/2$ and $f=1$). Measurements are carried out at 4.2 K.}
\label{dv}
\end{figure}

\begin{figure}
\includegraphics[width=0.45\textwidth]{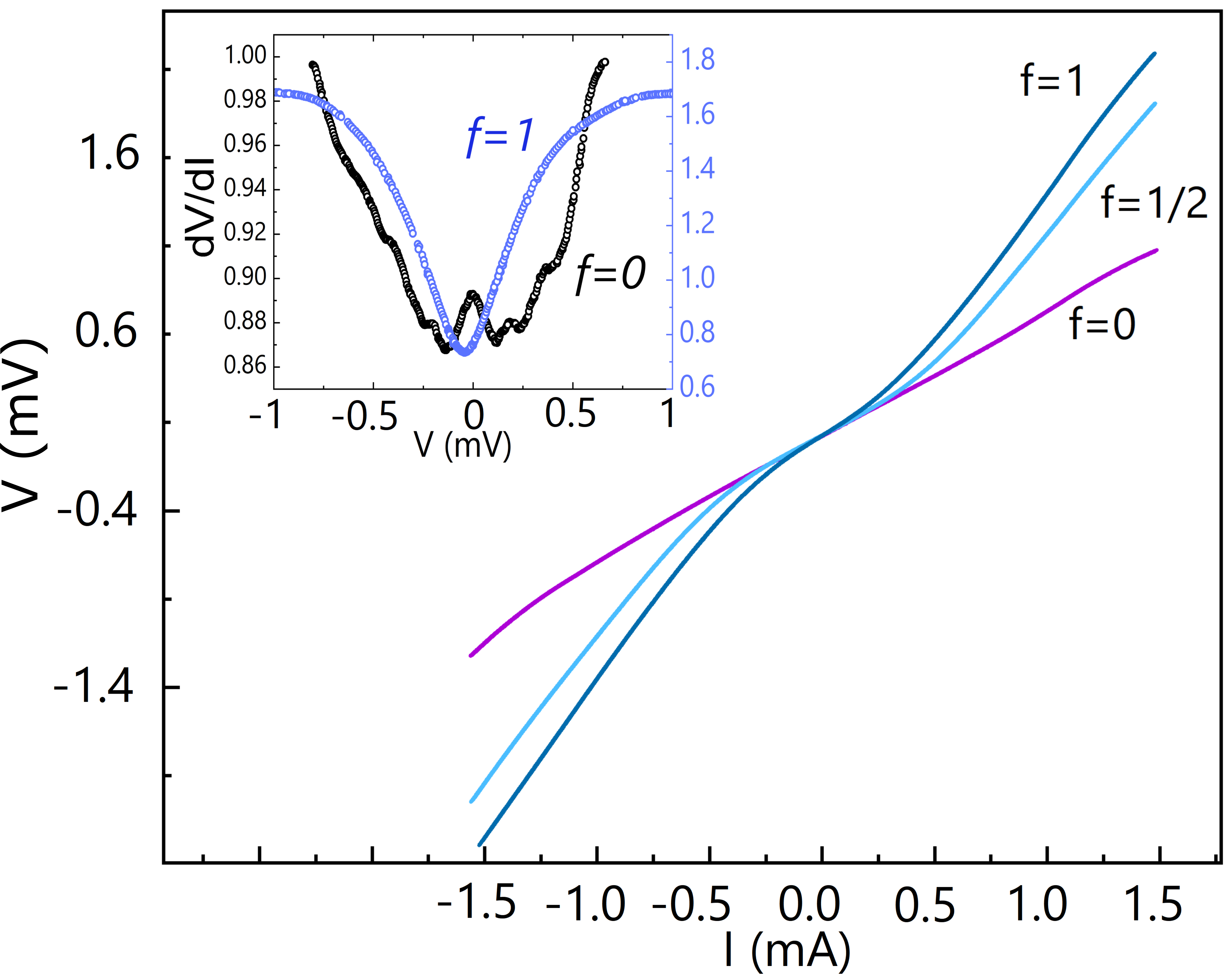}
\caption{Measured I-V curves at different applied external magnetic fields. The increase of the magnetic field triggers the non-linearity in the I-V nature. (Inset): The Andreev spectra derived from the numerical derivation of the I-V curves. The reflection peaks observable at zero field are shifted towards the higher potentials. Measurements are carried out at 4.2 K.}
\label{IV}
\end{figure}

The normalized resistance of the device measured versus temperature around the critical temperature is shown in figure \ref{Tc}. The proximity array device (PAD) shows a multi-stage transition around a critical temperature T$_c$. It can be understood considering the coherence extension via decrease of the temperature among the coupled superconducting islands, each of which composed of grains. At T$\geq$T$_1$ the grains on each islands have incoherent non-superconducting phase (i.e., normal state). Reaching T$_1$ the phase coherence develops between the islands’ grains and the system resistance decreases. 
Below T$_1$, the intra-island phase coherence strengthens continuously and the system resistance continuously decreases rather than steeply dropping at T$_1$. In this region the normal-metal coherence length increases until it becomes comparable to the island spacing. Then, inter-island phase coherence begins to emerge and finally reaching T$_2$ the system undergoes a transition to a fully superconducting state \cite{Resnick1981,Abraham1982,Pinto2018}.
The dynamic resistance of the PAD was measured at 4.2 K, being in the fully superconducting state, as a function of the applied magnetic field and current (see Figure \ref{dv}). The minima reflect the modulation of the ground-state energy due to formation of periodic vortex patterns in a magnetic field. The presence of these dips in the resistance and singularities in magnetization were observed in previous experiments \cite{Baturina2011}. However, a reversal of the minima in dV/dI into maxima by increasing the current bias in our case can be interpreted as a direct manifestation of the existence of the vortex Mott insulator and its transition into a metallic state. This transition occurs specifically at integer and fractional frustration fields $f$ being the $B/B_0$ with $B_0=27.8$ mT, the ratio of the magnetic quantum flux ($\Phi_0$) on the area of the junction. As expected, the transitions are most pronounced at integer and half-integer frustration factors ($f=1/2$ and $f=1$) \cite{Poccia2015}. At applied currents above I=2.6 mA, the emergence of a dissipative regime is clearly observable in which the singularities tends to be wiped out. Since the resistance of our device shows pronounced dips even at large currents (not shown),  indicating strong pinning, neither the minima-to-maxima flips nor the dissipative regime cannot be explained as vortex depinning \cite{Poccia2015,Benz1990}. Hence, it can be deduced that at such high current values, the metallic state features are diminished by the contribution of quasi particles current.  
On the other hand, the I-V measurements at different applied magnetic field (see figure \ref{IV}) gradually shows non linearity. At the highest applied magnetic field in our experiment the I-V curve tends to exhibit a $S$ shape behavior. This indicates a super-relativistic motion of non-localized vortices, in which they can overcome the characteristic electromagnetic wave velocity \cite{Zitzmann2002}. This motion should be accompanied by Cherenkov radiation of Josephson plasma waves with a dispersion relation of the eigen modes given by $\omega_m^2=\omega_p^2+c_m^2k^2$ using the Sine-Gordon equations where $c_m$ are the characteristic velocities \cite{Sakai1993,Zitzmann2002,Goldobin1998}. However, in our proximity array the Cherenkov oscillations induced by moving fluxons are damped by the increased quasi-particle conductance. In consequence, the decay length of the oscillating tail created by the fluxon decreases and the resonances on the fluxon step ceases to exist that, in agreement with the observed $S$ shape behavior. The observation of the resonance steps can be expected either via an increase of the magnetic field (below the H$_{c2}$) or by the decrease of the measurement temperature to values well below the T$_2$.   

The representative of the Andreev spectra (shown in figure \ref{IV} (inset)) in our PAD was achieved by numerical derivation of the I-V curves. The reflection peaks at zero field spectra is clearly visible from which a gap value of the PAD can be obtained as $\Delta\sim$1.3 meV. At the magnetic field corresponding to the frustration $f=1$, the gap peculiarities in the spectra are not only shifted towards lower energies as it is expected in the classical case, but they move in the inverse direction and their intensity is enhanced. This anomalous behavior implies that in non-localized vortex (metallic state) a triplet Cooper pairing with the total spin moment of a superconducting $S=1$ may occur. In this case a moderate magnetic field can contribute to the stabilization or even enhancement of the superconducting states \cite{Tran2012} and hence the superconducting order parameter. 
\section{Conclusions}
In this contribution we show the vortex dynamics in proximity array device (PAD) vs. external parameters (e.g., magnetic field and current) that can result in a transition of the vortices from a localized (Mott insulating) to a non-localized metallic state. We have also demonstrated a super relativistic motion in the non-localized metallic regime for which the Josephson plasma frequency can be tuned. The application of the magnetic field also enhances the superconducting order parameter and hence the superconducting gap can be modulated via the external magnetic field. Furthermore, in non-equilibrium superconductivity theory, the presence of Cooper pairs in the normal metal is described by the anomalous Green function $F^R(x,\epsilon)= -isin \Theta(x, \epsilon)$, expressed by the proximity angle $\Theta (x,\epsilon)$ which is a complex function of both the energy and the position \cite{Larkin1975}. The Josephson plasma frequency in the super-relativistic motion of the fluxons can also be tuned via the matrix parameter $k$. Hence, further modulations can be achieved by pre-designed Nb island geometries to meet the required superconducting properties. Our results suggest that the superconducting properties (e.g., supeconducting gap) and subsequently the photo-response of the dynamic array of the proximity junctions can be extensively tuned via the external parameters i.e., Geometry, Magnetic and electric field and Matrix parameters \cite{Halsey1991}. The tunability of such dynamic array can be exploited for sensitive radiation detection in a broad band, particularly in Terahertz region. 

\section{Acknowledgment}
This research has been mainly performed within  the framework of the TERA project, a proposal funded by the INFN V$^{th}$ Committee.
\bibliography{ACTA-PAD}
\end{document}